\documentclass[referee]{raa}            

\usepackage{graphicx,times}             

\begin{document}

   \title{A magnetic model for low/hard state of black hole binaries
$^*$
\footnotetext{$*$ Supported by the National Natural Science Foundation of China.}
}

   \volnopage{Vol.0 (2015) No.0, 000--000}      
   \setcounter{page}{1}          

   \author{Yong-Chun Ye
      \inst{1}
   \and Ding-Xiong Wang
      \inst{1}
   \and Chang-Yin Huang
      \inst{2}
   \and Xiao-Feng Cao
      \inst{3}
   }

   \institute{School of Physics, Huazhong University of Science and Technology, Wuhan,430074,China; {\it dxwang@hust.edu.cn}\\
        \and
             School of Physics and Optoelectronic Engineering, Yangtze University, 434023, Jingzhou, China\\
        \and
             School of Physics and Electronics Information, Hubei University of Education, 430205 Wuhan, China\\
   }

   \date{Received~~2015 month day; accepted~~2015~~month day}

\abstract{ A magnetic model for low/hard state (LHS) of black hole X-ray binaries (BHXBs), H1743-322 and GX 339-4, is proposed based on the transportation of magnetic field from a companion into an accretion disk around a black hole (BH). This model consists of a truncated thin disk with an inner advection-dominated accretion flow (ADAF). The spectral profiles of the sources are fitted in agreement with the data observed at four different dates corresponding to the rising phase of the LHS. In addition, the association of the LHS with quasi-steady jet is modelled based on transportation of magnetic field, where the Blandford-Znajek (BZ) and Blandford-Payne (BP) processes are invoked to drive the jets from BH and inner ADAF. It turns out that the steep radio/X-ray correlations observed in H1743-322 and GX 339-4 can be interpreted based on our model. 
}

   \authorrunning{Y.-C. Ye et al. }            
   \titlerunning{A magnetic model for low/hard state of black hole binaries}  

   \maketitle

\section{Introduction}           
\label{sect:intro}

Although a consensus on classification of spectral states of black hole X-ray binaries (BHXBs) has not been reached, it is widely accepted that these states can be reduced to only two basic ones, i.e., a hard state and a soft state, and jets can be observed in the hard state, but cannot be in the soft one. The accretion flow in low/hard state (LHS) is usually supposed to be related to a truncated thin disk with an inner advection-dominated accretion flow (ADAF) in the prevailing scenario (Esin et al. 1997, 1998, 2001; McClintock {\&} Remillard 2006; Done, Gierlinski {\&} Kubota 2007; Yuan {\&} Narayan 2014 and references therein). Generally speaking, the thermal component of the spectra of BHXBs can be well fitted by the outer thin disc, while the power-law component can be interpreted by the inner ADAF. Although the ADAF model is rather successful in interpreting spectral state of BHXBs, it has important limitations as pointed out by some authors (e.g., McClintock {\&} Remillard 2006). Besides the power-law component dominates, another feature of LHS of BHXBs is its association with quasi-steady jets. Although ADAF model is successful in fitting the spectra of LHS of some BHXBs, the detail of how the associated jets are produced has not been addressed. Another feature of LHS is that the universal radio--X-ray correlation has been found for a sample of BHXBs (Hannikainen et al. 1998; Corbel et al. 2003; Gallo, Fender {\&} Pooley 2003; Wu {\&} Cao 2006; Wu {\&} Gu 2008; Corbel et al. 2013). However, more and more ¡®outliers¡¯ were found in last few years, which evidently deviate from the universal radio--X-ray correlation and usually show a much steeper correlation with an index of $\sim1.4$ (e.g.,H1743-322, Jonker et al. 2010; Coriat et al. 2011; Swift 1753.5-0127, CadolleBel et al. 2007; Soleri et al. 2010; XTE J1650-500, Corbel et al. 2004; XTE J1752-223, Ratti et al. 2012).

On the other hand, the most remarkable feature of the state transitions of BHXBs is a phenomenon known as hysteresis, behaving a counterclockwise q-shaped curve in hardness-intensity diagram (HID) (Miyamoto et al. 1995; Homan et al. 2001; Fender, Belloni {\&} Gallo 2004; Belloni et al. 2005; Homan {\&} Belloni 2005; Fender et al. 2009; Fender {\&} Belloni 2012).However, the hysteresis effect cannot be interpreted by only accretion rate in ¡®one-dimensional¡¯ picture of the ADAF model (see Yu {\&} Yan 2009; Kylafis {\&} Belloni 2015).

There an exception in Cyg X-1 based on the continuous monitoring of Cyg X-1 in the 1.3--200 keV band by using All-Sky Monitor/ Rossi X-Ray Timing Explorer and BATSE/Compton Gamma Ray Observatory for about 200 days from 1996 February 21 to early September. It is found that the luminosity of the spectral state transitions in Cyg X-1 is quite constant (Zhang et al. 1997, hereafter ZCH97), thus no q-shaped curve would be seen in HID.

It is noted that two important discoveries are related closely to the state transitions of X-ray binaries: (i) the correlation between the transition luminosity in the rising phase and the outburst amplitude, which demonstrated that the scale of hysteresis is not arbitrary (Yu et al. 2004); and (ii) the correlation between the transition luminosity of the state transition on the rising phase and the rate-of-increase of the X-ray luminosity, which supports that the rate-of-change of the mass accretion rate is the apparent second governing parameter (Yu {\&} Yan 2009).

In this paper, we intend to model the LHS of two BHXBs, H1743-322 and GX 339-4, based on the transportation of the magnetic field into the inner ADAF via the outer thin disc, where the magnetic field is carried by the plasma from the companion. Two promising mechanisms for powering jet, i.e., the Blandford-Znajek (BZ, Blandford {\&} Znajek 1977) and Blandford-Payne (BP, Blandford {\&} Payne 1982) processes are invoked for interpreting the association of a quasi-steady jet with LHS. It turns out that the spectral profiles of the both BHXBs are fitted in agreement with the data observed at four different dates in the rising phase of the LHS, and the association of LHS with a quasi-steady jet is satisfied with the steep radio/X-ray correlations for these two sources.

This paper is organized as follows. In Section 2 we give a brief description of our model, and a scenario of magnetic field transported with the accreting plasma from the companion is proposed. It is assumed that both accretion rate and magnetic field increase in LHS according to power-law of different indexes. In Section 3 the spectral profiles of LHS of H1743-322 and GX 339-4 are fitted based on the data observed in four different date in 2003 and 2010, respectively. In Section 4 we fit the radio--X-ray correlation for H1743-322 and GX 339-4, both behaving as ¡®outliers¡¯ in a power-law with much steeper index than the universal correlation. Finally in Section 5, we discuss some issues of the role of magnetic field in state transition of BHXBs.

\section{DESCRIPTION OF OUR MODEL}

It is widely believed that the most promising mechanisms for powering jets are the BZ and BP processes, which rely on a poloidal, large-scale magnetic field anchored on a spinning black hole (BH) and its surrounding accretion disk (Blandford {\&} Znajek 1977; Blandford {\&} Payne 1982; Livio 2002; Lei et al. 2005; Lei et al. 2008; Wu, Cao {\&} Wang 2011; Doeleman et al. 2012; Wu et al. 2013; for a review see Spruit, 2010).Recently, a number of authors discussed the role of the magnetic field in state transition of BHXBs (Cao 2011; Miller et al. 2012; King et al. 2012; Sikora {\&} Begelman 2013; Dexter et al. 2014). However the origin of the magnetic field in BHXBs remains unclear.

In this paper, we propose a scenario of evolution of magnetic field in the outbursts of BHXBs based on the transportation of magnetic field into accretion disk from the companions. The physical picture of the magnetic field transportation is depicted in Fig. 1, and the main features are given as follows.

(i) The magnetic field is carried by the plasma from the companions, which is transported to an accretion disk around a spinning BH.

(ii) The accretion disk consists of two parts, i.e., an inner advection-dominated accretion flow (ADAF) and an outer thin disc, between which the truncated radius is $R_{tr}$.

(iii) The relation between large-scale and tangled small-scale magnetic field on the disk is given as follows (Livio, Ogilvie {\&} Pringle 1999).


\begin{equation}
\label{eq1} B_L \sim (H/R) B_S ,
\end{equation}

\begin{figure}
\vspace{0.5cm}
\begin{center}
\includegraphics[width=8cm]{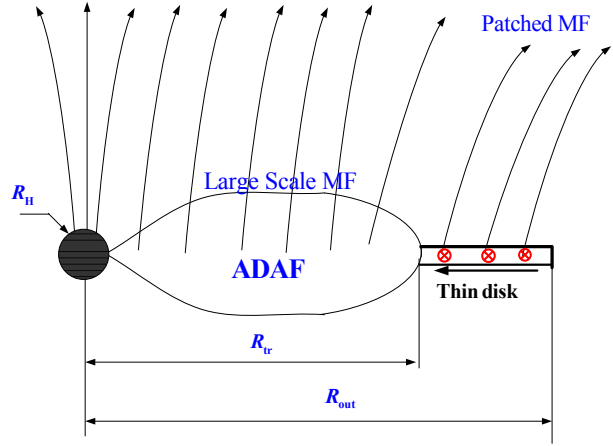}
\caption{The large scale magnetic field transported to the inner ADAF with the patched fields (represented by the symbol '$\otimes$' ) transported through the outer thin disk, being apt to overcome the magnetic diffusion. The truncated and the outer radii of the thin disk are represented by $R_{tr}$ and $R_{out}$, respectively.} \label{fig1}
\end{center}
\end{figure}

\noindent where $H$ and $R$ are the half height and disk radius, and $B_S$ and $B_L$  are small- and large-scale magnetic fields on the disc, respectively.

In addition, four assumptions for the transportation of large scale magnetic field are given as follows.

(i) The patched magnetic fields are assumed in the outer thin disk due to MHD turbulence, reducing the outward magnetic diffusion in the thin disk significantly as argued by Spruit {\&} Uzdensky (2005), thus the magnetic field can be transported across the truncated radius.

(ii)The vertical component of the large scale poloidal field keeps the same direction in the inner ADAF during the rising phase of the LHS, as a number of authors did (Lubow et al. 1994; Spruit {\&} Uzdensky 2005; Igumenshchev 2009; Cao 2011; Guilet {\&} Ogilvie 2012; Sikora {\&} Begelman 2013).

(iii)The strength of the magnetic field at the BH horizon is related to that at the inner ADAF by $B_{H} = 50 B_{ADAF}$ based on the results given by Cao (2011), where $R_H$ represents the radius of the BH horizon, and $B_H$ and $B_{ADAF}$ are the magnetic fields on the BH horizon and ADAF, respectively.

(iv) Jet contribution to the X-ray luminosity is distributed uniformly to the whole X-ray band of the spectra of LHS due to lack of detailed origin of the jet radiation.

Thus the magnetic flux can be carried by the accreting plasma to the ADAF through the truncated thin disc, and the accumulated flux in ADAF is related to the accretion rate as follows.


\begin{equation}
\label{eq2} \Phi ( t )  = \int_{t_b}^{t} {\dot{\Phi} dt} = 2^{3/2} \pi \delta^{-1/2} \beta^{1/2} \dot{M}_{Edd}^{1/2} \int_{t_b}^{t} {\lambda_m^{1/2} (\dot{m}c)^{1/2}c dt} .
\end{equation}

The quantity $\Phi ( t )$ in equation (2) is the magnetic flux carried by the accreting plasma, and it can be calculated by integrating from $t_b$ to $t$, which correspond to the beginning of LHS and the subsequent rising phase of LHS, respectively (Fender, Belloni {\&} Gallo 2004; Fender {\&} Belloni 2012).

The ratio of magnetic energy density to mass energy density of accreting plasma is defined as $\lambda_m \equiv B^2 / (8 \pi \rho c^2)$, and $\beta = v_{tr} / c$ is the ratio of radial velocity of the accreting plasma at the truncated radius to speed of light, and $\delta = H_{tr} / R_{tr}$  is the ratio of half height to disk radius at the truncated radius. The accretion rate is defined in terms of Eddington accretion rate, $\dot{m} \equiv \dot{M} / \dot{M}_{Edd}$, which is related to Eddington luminosity by


\begin{equation}
\label{eq3} \dot{M}_{Edd} = L_{Edd} / 0.1 c^2 = 1.4 \times 10^{18} m_{_H} (g \cdot s^{-1}),
\end{equation}

\noindent where $m_{_H} = M_H / M_{\odot}$ is the BH mass in terms of solar mass.

Equation (2) can be easily derived based on the the picture of magnetic flux transportation. From Fig. 1 and equation (2) we find that the magnetic flux within the truncated radius $R_{tr}$ can be estimated by


\begin{equation}
\label{eq4} \begin{array}{l} \Phi_{tr} = \pi (R_{tr}^2 - R_{H}^2) B_{ADAF} + 2 \pi R_H^2 B_H
\\
\\
\quad\quad= 0.02 \pi B_H (R_{tr}^2 + 99R_H^2),
\end{array}
\end{equation}

\noindent where $R_H$ represents the radius of the BH horizon, and $B_H$ and $B_{ADAF}$ are the magnetic fields on the BH horizon and ADAF, respectively. In equation (4), $B_H = 50 B_{ADAF}$ is assumed based on the results given by Cao (2011).

In order to discuss the evolution of the magnetic field in LHS we assume that the accretion rate at the truncated radius $R_{tr}$ and the magnetic field at the BH horizon can be expressed as the power functions of the outburst time $t$, i.e.,


\begin{equation}
\label{eq5} \dot{m}_{_{R_{tr}}} = \dot{m}_0 (t / \tau)^{\alpha_m},
\end{equation}


\begin{equation}
\label{eq6} B_H = B_0 (t / \tau)^{\alpha_{_B}}.
\end{equation}

\noindent where $\dot{m}_{_{R_{tr}}}$ and $B_H$ represent the accretion rate at the truncated radius $R_{tr}$ and the magnetic field at the BH horizon, respectively.

The parameter $\tau$ in equations (5) and (6) is the duration between $t_b$ and the time reaching the intermediate state (IMS), going through a rising phase of LHS. We assume $ 0 \leq t / \tau \leq 1 $ by considering that both $\dot{m}$ and $B_H$ attain their maxima $\dot{m}_0$ and $B_0$ in IMS, respectively. The value of $\tau$ varies from several weeks to several months for different BHXBs in different outbursts. Incorporating equations(4)--(6), we have $R_{tr}$ given by


\begin{equation}
\label{eq7} R_{tr}^2 = 100\sqrt{2} \delta^{-1/2} \beta^{1/2} c^{3/2} \dot{M}_{Edd}^{1/2} \tau \frac{\lambda_m^{1/2}(t/\tau)^{\alpha_{tr}}}{B_0[(\alpha_m / 2)+1]} - 99 R_H^2,
\end{equation}

\noindent where


\begin{equation}
\label{eq8} \alpha_{tr} = (\alpha_m / 2) - \alpha_{_B} + 1
\end{equation}

It is found from equations (7) and (8) that $\alpha_{tr}$ is related to $\alpha_m$ and $\alpha_{_B}$, and the truncated radius $R_{tr}$ increases/decreases with time for positive/negative $\alpha_{tr}$, provided that the values of $\beta$, $\delta$, $B_0$, $\lambda_m$ and $\tau$ are fixed.

Some authors proposed an excellent description of the general picture of spectral evolutions of BHXBs with the Hardness-Intensity-Diagram (HID). As shown in HID that LHS is associated with a steady jet with increasing X-ray luminosity, while hardness remains almost unchanged(Fender, Belloni {\&} Gallo 2004; Fender {\&} Belloni 2012). As mentioned above, in our model the accretion flow consists of an inner ADAF and an outer thin disc. In order to interpret the association of LHS with quasi-steady jets, we invoke the two most promising mechanisms for powering jets, i.e., the BZ and BP processes, which are related closely to the large scale magnetic fields in the accretion process.

During the past decade both numerical simulations (Stone, Pringle {\&} Begelman 1999; Hawley {\&} Balbus 2002; Igumenshchev, Narayan {\&} Abramowicz 2003) and analytical calculations (Narayan {\&} Yi 1994, 1995; Blandford {\&} Begelman 1999; Narayan, Igumenshchev {\&} Abramowicz 2000; Quataert {\&} Gruzinov 2000) indicate that only a fraction of the plasma accretes onto the BH and the rest is ejected from the outflow. We calculate the global solution of the accretion flow based on ADAF with truncated thin disk (e.g., Yuan 2001; Yuan et al. 2005). And then, we solve the radiation hydrodynamics equations self-consistently, obtaining the advection factor $f(r)$ as the function of the radius, i.e.,


\begin{equation}
\label{eq9} f(r) = q_{adv} / q_{vis} = (q_{vis} - q_{ie}) / q_{vis}
\end{equation}

\noindent where $q_{adv}$, $q_{vis}$ and $q_{ie}$ are the rates of energy advection, viscous heating, and Coulomb collision cooling for the ions, respectively. Meanwhile, Comptonization is treated as a local approximation. The following radiation processes, i.e., bremsstrahlung, synchrotron emission, and the Comptonization of both synchrotron photons from the hot accretion flow and soft photons from the cool disk outside transition radius are included. The emission from the outer cool disk is modeled as a multicolor blackbody spectrum. The effective temperature as a function of radius is determined by the viscous dissipation and the irradiation of the disk by the inner hot ADAF.

Here, the concerning parameters of the accretion flow are the truncated radius $R_{tr}$, the accretion rate $\dot{m}$, and temperature $T_{tr}$ of the accretion flow at $R_{tr}$ is regarded as an outer boundary condition of the ADAF region(Yuan 1999). In our model, $\dot{m}$ and $R_{tr}$ can be calculated by using equations (5) and (7), and $T_{tr} = 10^9 K$ is taken in a simplified analysis. Furthermore, we take $\alpha = 0.3$ as the ¡°typical¡± values of the viscosity parameter, and assume that $\sim10 {\%} $ of the viscous dissipation heats electrons directly. So, in our calculations, the free parameters are  $\alpha_m$, $\alpha_{_B}$ and $\beta_{gas}$, which is defined as the ratio of gas pressure to the sum of gas pressure and magnetic pressure. It is found in calculations that $\beta_{gas}$ decreases continuously with time due to the transportation of the magnetic field into the inner ADAF.

A number of parameters are involved in our model, being classified into two types based on their roles in the fittings, and a summary is given as follows.

Type I: Six free parameters are assigned fixed values before fittings, i.e., $\dot{m}$, $B_0$, $\lambda_m$, $\alpha_m$, $\beta$, $\delta$ as indicated in section 3.

Type II: The values of five parameters are determined in the fittings. Specifically, $t / \tau$, $\dot{m}$, $\alpha_{_B}$ and $\beta_{gas}$ are determined in fitting the spectra of H1743-322 and GX 339-4 as shown in Tables 1 and 2 respectively, while those of $\eta_i$ are determined in fitting the steep radio/X-ray correlations of these two BHXBs by combining the jet contributions on the $L_R$ and $L_X$ as given in section 4.

All the fittings for H1743-322 and GX 339-4 are based on the data observed in four different date in 2003 and 2010, respectively.

\section{FITTING SPECTRAL PROFILES OF LHS OF THE BHXBS}

\subsection{H1743-322}

Now, we fit the spectra of the rising phase of LHS of H1743-322 based on four different observation data, which consists of two steps. First, we extract four energy spectra of H1743-322 with observations obtained by RXTE PCA(Obs. ID: 80138-01-01-00; Obs. ID: 80138-01-02-00); Obs. ID: 80138-01-03-00 and Obs. ID: 80138-01-05-00), and set the rising phase of LHS duration of H1743-322 as  $\tau = 25 day = 2.45 \times 10^6 sec$ (Joinet et al. 2005; Remillard et al. 2006), and then we can get the value $t / \tau$ of each date in LHS. Second, combining the equations (5) and (7), we calculate  $\dot{m}$ and $R_{tr}$, and fit the LHS spectra of H1743-322 based on the truncated disk model, where $\dot{m}_0 =0.03$, $\lambda_m = 10^{-24}$,  $\beta = 0.5$ and $\delta = 1$ are adopted. The BH mass is taken as $M = 10 M_{\odot}$, the distance to the source as $D = 8 kpc$ , and the binary inclination as $ \theta = 60^{\circ}$(McClintock, Remillard {\&} Rupen 2009; Blum et al. 2010; Motta, Munoz-Darias {\&} Belloni 2011; Coriat et al. 2011). Following Esin (2001) and Yuan, Cui {\&} Narayan (2005), and invoking the formula given by Paczynski (1971), we estimate the outer radius of the truncated thin disk as $R_{out} = 3\times 10^4 R_H (10 M_{\odot} / M)^{2/3}$.

\begin{figure}
\vspace{0.5cm}
\begin{center}
{\includegraphics[width=6.1cm]{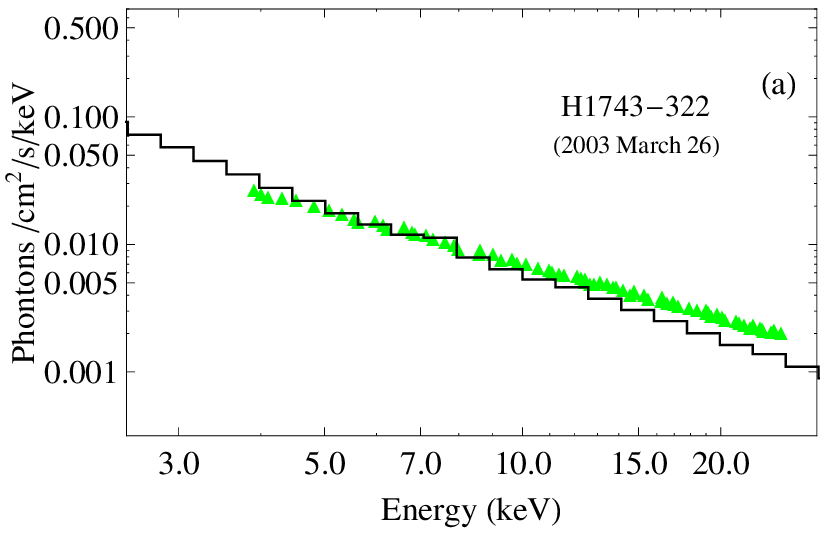} \hfill
\includegraphics[width=6.1cm]{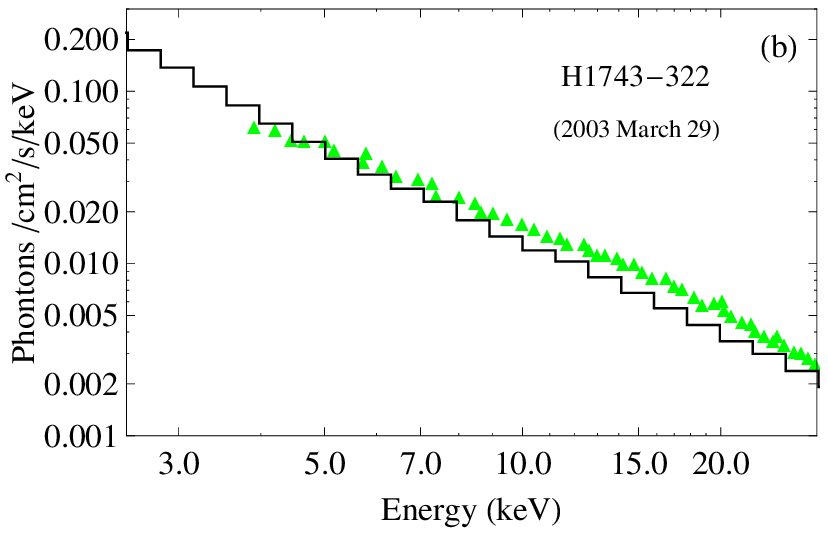} \hfill
\includegraphics[width=6.1cm]{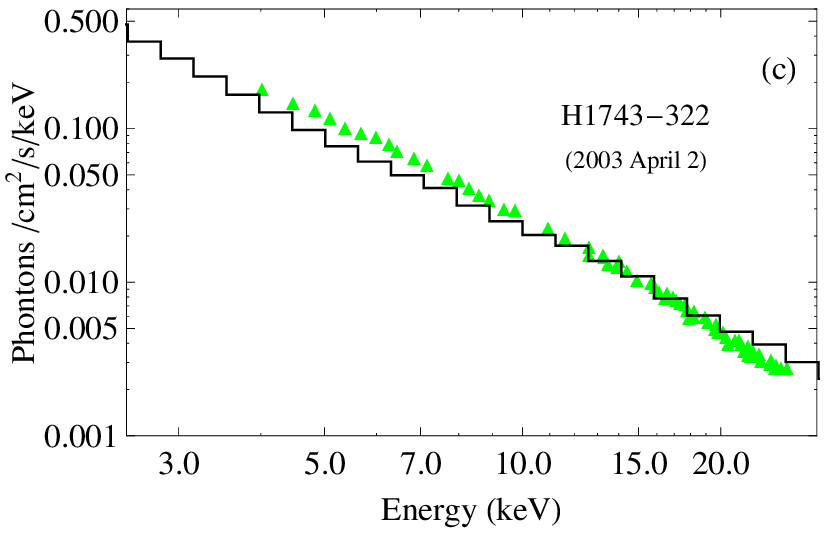} \hfill
\includegraphics[width=6.1cm]{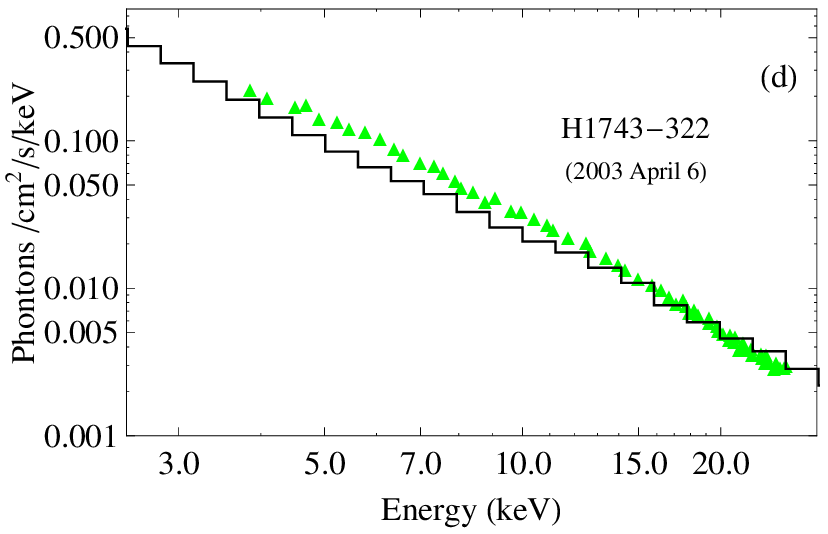}
 }
\caption{The spectra of LHS of H1743-322 in the jagged lines fitted to the green triangles based on the four different dates: (a) 2003 March 26 (Obs. ID: 80138-01-01-00); (b) 2003 March 29 (Obs. ID: 80138-01-02-00); (c) 2003 April 2 (Obs. ID: 80138-01-03-00) and (d) 2003 April 6 (Obs. ID: 80138-01-05-00).} \label{fig2}
\end{center}
\end{figure}

\begin{table*}
\caption{The parameters for fitting spectra of LHS of H1743-322 in four different dates}
\begin{tabular}
{|p{342pt}|p{47pt}|} \hline
Input  &  Output
\end{tabular}
\begin{tabular}
{|p{47pt}|p{47pt}|p{47pt}|p{47pt}|p{47pt}|p{47pt}|p{47pt}|} \hline Date & $t / \tau$ & $\dot{m}$ & $\alpha_m$ & $\alpha_{_B}$ & $\beta_{gas}$ & $r_{tr}$\\
\hline 1& 0.01& 0.003 & 0.5 & 0.3 & 0.86 & 200 \\
\hline 2& 0.13& 0.011 & 0.5 & 1.3 & 0.82 & 123.89\\
\hline 3& 0.29& 0.016 & 0.5 & 1.3 & 0.80 & 121.43\\
\hline 4& 0.45& 0.020 & 0.5 & 1.3 & 0.77 & 120.10\\
\hline
\end{tabular}
 \label{tab1}
 \begin{minipage}{140mm}
Note: Date 1, 2, 3 and 4 in Table 1 indicate March 26, March 29, April 2 and April 6 in 2003, respectively. The dimensionless truncated disk radius is defined as $r_{tr} = R_{tr} / R_H$.
 \end{minipage}
\end{table*}

\begin{figure}
\vspace{0.5cm}
\begin{center}
{\includegraphics[width=6.1cm]{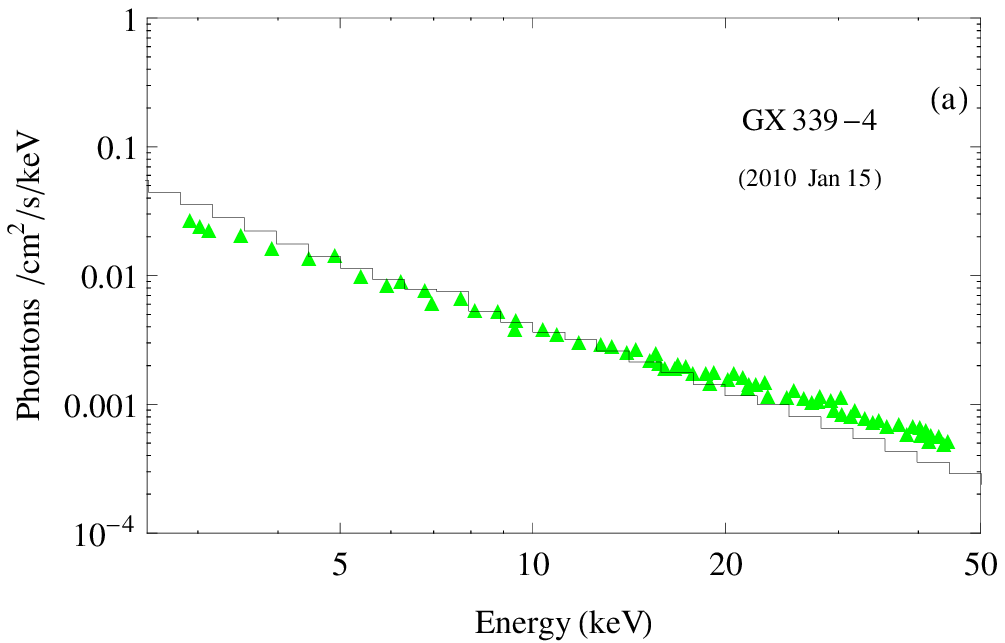} \hfill
\includegraphics[width=6.1cm]{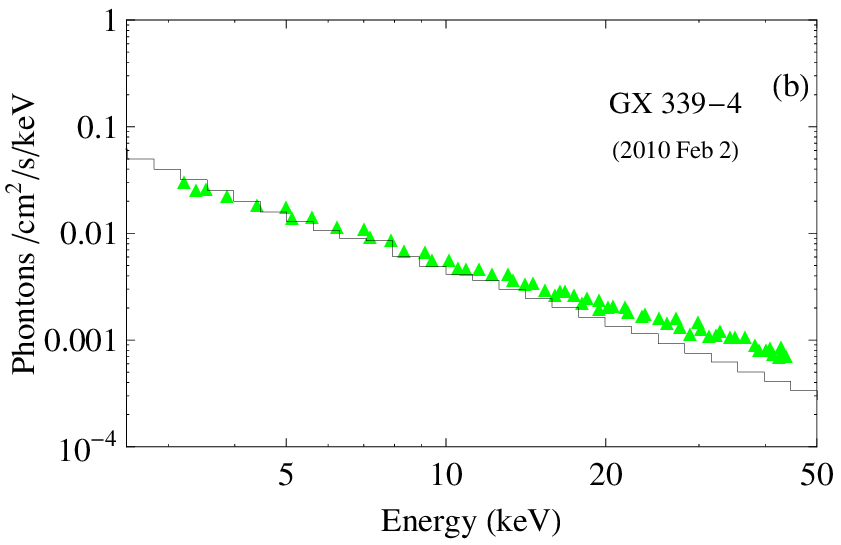} \hfill
\includegraphics[width=6.1cm]{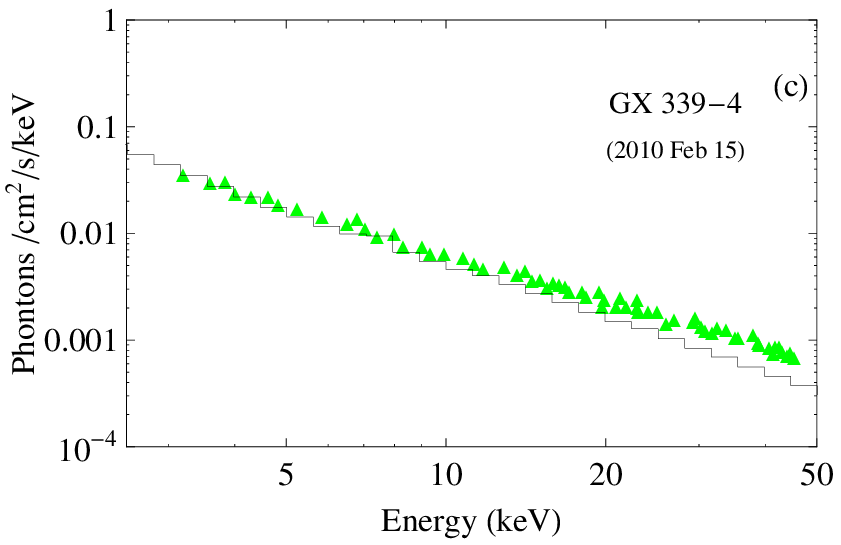} \hfill
\includegraphics[width=6.1cm]{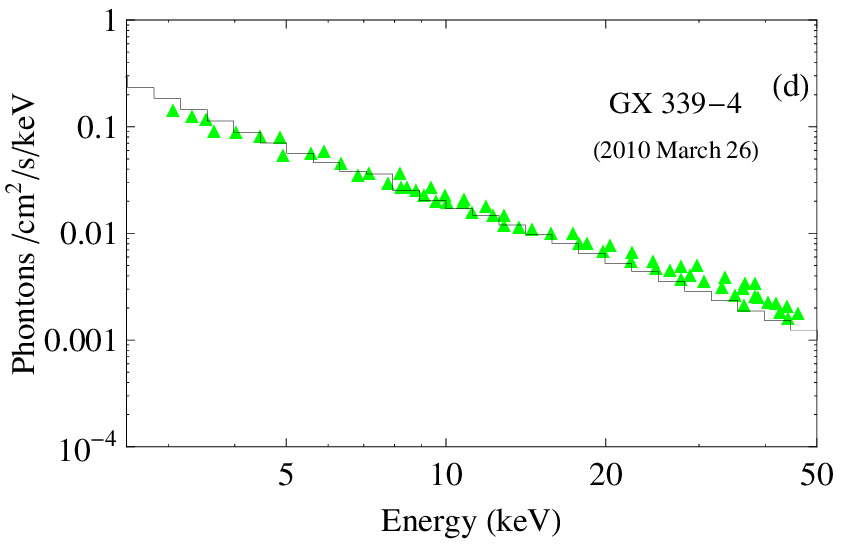}
 }
\caption{The spectra of LHS of GX 339-4 in the jagged lines fitted to the green triangles based on the four different dates: (a) 2010 Jan 15 (Obs. Id: 95409-01-02-02); (b) 2010 Feb 2 (Obs. ID: 95409-01-04-02); (c) 2010 Feb 15 (Obs. ID: 95409-01-06-01) and (d) 2010 March 26 (Obs. Id: 95409-01-12-00).} \label{fig3}
\end{center}
\end{figure}

\begin{table*}
\caption{The parameters for fitting spectra of LHS of GX 339-4 in four different dates}
\begin{tabular}
{|p{342pt}|p{47pt}|} \hline
Input  &  Output
\end{tabular}
\begin{tabular}
{|p{47pt}|p{47pt}|p{47pt}|p{47pt}|p{47pt}|p{47pt}|p{47pt}|} \hline Date & $t / \tau$ & $\dot{m}$ &  $\alpha_m$ & $\alpha_{_B}$ & $\beta_{gas}$ & $r_{tr}$\\
\hline 1& 0.01& 0.003 & 0.5 & 0.3 & 0.87 & 300 \\
\hline 2& 0.12& 0.010 & 0.5 & 1.3 & 0.85 & 250.84 \\
\hline 3& 0.24& 0.015 & 0.5 & 1.3 & 0.84 & 246.56 \\
\hline 4& 0.59& 0.023 & 0.5 & 1.3 & 0.81 & 241.00 \\
\hline
\end{tabular}
 \label{tab2}
 \begin{minipage}{170mm}
Note: Date 1, 2, 3 and 4 in Table 2 indicate Jan. 15, Feb. 2, Feb. 15 and March 26 in 2010, respectively.
 \end{minipage}
\end{table*}

The fitted spectra of LHS of H1743-322 and the related free parameters are given in Fig. 2 and Table 1, respectively. Some authors pointed out that there exists a truncated disk in H1743-322 (e.g., Sriram, Agrawal {\&} Rao 2009), and our fitting results of the radius $r_{tr}$ is consistent with their conclusion.

\subsection{GX 339-4}

The fitting steps for GX 339-4 are the same as those for H1743-322. For this source we take BH mass as $M = 5.8 M_{\odot}$(Hynes et al. 2003; Munoz-Darias,Casares {\&} Martinez-Pais 2008), and the distance to the source as  $D = 8 kpc$(Zdziarski et al. 2004; Hynes et al. 2004), and the binary inclination as  $ \theta = 30^{\circ}$(Cowley et al. 2002; Gallo et al. 2004; Miller et al. 2004, 2006, 2009; Reis et al. 2008; Done {\&} Diaz Trigo 2010), and the rising phase of LHS duration of GX 339-4 as  $\tau = 110 day = 1 \times 10^7 sec$(Nandi et al. 2012, Allured et al. 2013). The fitting results and the related parameters are shown in Fig. 3 and Table 2, respectively.

Our results show that it is reasonable to assume a truncated disk in GX 339-4. The strongest case for disk truncation was presented by Tomsick et al. (2009), in which fluorescent iron emission lines are invoked to find the inner accretion disk edge, being greater than Schwarzschild radius for GX 339-4 in LHS. Claims of a truncated disk in LHS also appear based on modeling of direct disk emission accounting for irradiation of the inner disk (Cabanac et al. 2009). Recently, Plant et al.(2014) pointed out that there exists a truncated disk in GX 339-4, and it decreases monotonically with time. As shown in Table 2, the variation of the truncated disk of GX 339-4 obtained in our model is consistent with the conclusion given by Plant et al. (2014).

As shown in Table 1 and Table 2, the truncated radii $R_{tr}$ move inward in the rising phase of LHS from 200 to 120.10 and from 300 to 241 corresponding to the variation of accretion rates from 0.003 to 0.020 and from 0.003 to 0.023 for H1743-322 and GX 339-4, respectively. This implies that the large-scale magnetic field is more and more concentrated in the inner ADAF in the rising phase of LHS due to the transportation of the magnetic field carried by the accreting plasma. It is also noted that these values of accretion rates are much less than the critical values, $\sim 0.35-0.4$, guaranteeing the validity of ADAF model in fitting the LHS of these two sources.

\begin{table*}
\caption{The parameters for fitting the correlation between $L_R$ and $L_X$ for H1743-322 and GX 339-4 based on four different observations}
\begin{tabular}
{|p{90pt}|p{90pt}|p{90pt}|p{90pt}|} \hline Source & Spin & $f_0$ & $B_0$  (Gauss)  \\
\hline H1743-322 & 0.2& 0.33& $ 7.9 \times 10^8$  \\
\hline GX 339-4 & 0.5& 0.33& $ 7.9 \times 10^8$  \\
\hline
\end{tabular}
 \label{tab3}
 \begin{minipage}{140mm}
Note: The spin $a_{*} = 0.2$ for H1743-322 is taken from Narayan et al. (2012), while $a_{*} = 0.5$ for GX 339-4 is assumed due to lack of data.
 \end{minipage}
\end{table*}

\section{THE RADIO--X-RAY CORRELATION}

The relation between radio luminosity $L_R$ and X-ray luminosity $L_X$ in LHS of BHXBs has been discussed by a number of authors (Corbel {\&} Fender 2002; Gallo, Fender {\&} Pooley 2003; Fender, Gallo {\&} Jonker 2003)


\begin{equation}
\label{eq10} L_R \propto L_{X}^b,
\end{equation}

\noindent where $b \sim 0.7 $ for $L_X$ in the 3-9 keV range. However, it has been shown recently that the radio and X-ray emission of some BHXBs are strongly correlated at high luminosity in LHS with much steeper power-law, i.e., $b \sim 1.4 $ in equation (10), and these sources including H1743-322 and GX 339-4 are referred to as ¡®outliers¡¯ (Jonker et al. 2010; Coriat et al. 2011; Cao, Wu {\&} Dong 2014).

Very recently, Huang, Wu {\&} Wang (2014, hereafter HWW14) modeled the steeper radio--X-ray correlation with slopes 1.2 based on the radiatively efficient disc-corona with the hybrid jet model, in which the radio emission is attributed to the jet with the X-ray emission from the disk corona. Although the steep correlation is fitted to the observed ones in HWW14, the simulated X-ray emission increases more and more slowly and almost becomes saturated at high accretion rates, which may be due to that they did not consider the contribution of the jet to the X-ray radiation. Compared to HWW14, this model focuses in the transportation of the magnetic field from the companions in which more attention is paid on the effect of the magnetic field transportation on the steep correlation in the rising phase of LHS observed in H1743-322 and GX 339-4.

In this paper, the BZ and BP mechanisms are invoked to drive the jet power in LHS of BHXBs, being regarded as follows,


\begin{equation}
\label{eq11} L_J = P_{BZ} + P_{BP},
\end{equation}

\noindent where $L_J$, $P_{BZ}$ and $P_{BP}$ are the jet power, the BZ power and the BP power, respectively. The BZ power is driven by the spinning BH via the large scale magnetic field on the horizon, and it reads (Macdonald {\&} Thorne 1982; Ghosh {\&} Abramowicz 1997)


\begin{equation}
\label{eq12} P_{BZ} = \frac{c}{32} \omega_F^2 B_H^2 R_H^2 a_{*}^2,
\end{equation}

\noindent where $\omega_{_F}$ is a parameter related to the angular velocity of the BH and that of the field line on the BH horizon by $\omega_F^2 = \Omega_F(\Omega_H - \Omega_F) / \Omega_H^2$. The maximum BZ power $P_{BZ}^{max}$ is reached as $\Omega_F = \Omega_H / 2$, i.e.,


\begin{equation}
\label{eq13} P_{BZ}^{max} = \frac{c}{128} B_H^2 R_H^2 a_{*}^2 = \frac{c}{128} B_0^2 R_H^2 a_{*}^2 (t / \tau)^{2\alpha_{_B}},
\end{equation}

\noindent where equation (6) for the evolution of the magnetic field in LHS is used in the last step.

The BP power can be expressed based on our previous work (Li, Gan {\&} Wang 2010) as follows,


\begin{equation}
\label{eq14} P_{BP} = \int_{R_H}^{R_{tr}} S_E 4 \pi R dR,
\end{equation}

\noindent where $S_E = B_{ADAF}^2 \Omega_{ADAF}^2 R^2 / (4 \pi c)$ is the energy flux driven in the BP process, and $\Omega_{ADAF}$ is the angular velocity of ADAF, being related to Keplerian angular velocity by


\begin{equation}
\label{eq15} \Omega_{ADAF} = f \Omega_k,
\end{equation}

\noindent where $f$ is defined as (Narayan {\&} Yi 1994; Yuan, Ma {\&} Narayan 2008, hereafter YMN08)


\begin{equation}
\label{eq16}
f = \left\{
\begin{array}{rl}
f_0, \quad\quad\quad\quad for \quad R \geq 3 R_H,\\
f_0 3 (R - R_H) / 2 R, \quad\quad\quad\quad R < 3 R_H.
\end{array}
\right.
\end{equation}

As argued in YMN08, the value of $f_0$ is taken as 0.33 for all accretion rates as the viscous parameter $\alpha$ is large. Incorporating equations (6), (13) and (15), we have


\begin{table*}
\caption{The fractions of jet power converted to X-ray luminosity corresponding to different observations}
\begin{tabular}
{|p{188pt}|p{188pt}|} \hline
H1743-322  &  GX 339-4
\end{tabular}
\begin{tabular}
{|p{50pt}|p{34pt}|p{34pt}|p{34pt}|p{50pt}|p{34pt}|p{34pt}|p{34pt}|} \hline Date & $ t/\tau $ & $\dot{m}$ & $\eta_i$ & Date & $ t/\tau $ & $\dot{m}$ & $\eta_i$ \\
\hline Mar. 26, 2003 & 0.01& 0.003&  0.20 & Jan. 15, 2010 & 0.01 & 0.003 & 0.25 \\
\hline Mar. 29, 2003 & 0.13& 0.011&  0.23 & Feb. 2, 2010 & 0.12 & 0.010 & 0.35 \\
\hline Apr. 2, 2003 & 0.29& 0.016&  0.28 & Feb. 15, 2010 & 0.24 & 0.015 & 0.39 \\
\hline Apr. 6, 2003 & 0.45& 0.020&  0.36 & Mar. 26, 2010 & 0.59 & 0.023 & 0.42 \\
\hline
\end{tabular}
 \label{tab4}

\end{table*}


\begin{equation}
\label{eq17}\begin{array}{l}
 P_{BP} = \frac{1}{c} \int_{R_H}^{3R_H} B_{ADAF}^2 \Omega_K^2 (R - R_H)^2 (9/4) f_0^2 R dR
 \\
 \\
 \quad\quad\quad + \frac{1}{c} \int_{3R_H}^{R_tr} B_{ADAF}^2 \Omega_K^2 f_0^2 R^3 dR
 \\
 \\
= (2\times 10^{-4} c)(B_0 f_0 R_H)^2 [(R_{tr}/R_H)-1.95](t / \tau)^{2a_{_B}}.
\end{array}
\end{equation}

\begin{figure}
\vspace{0.5cm}
\begin{center}
\includegraphics[width=6.1cm]{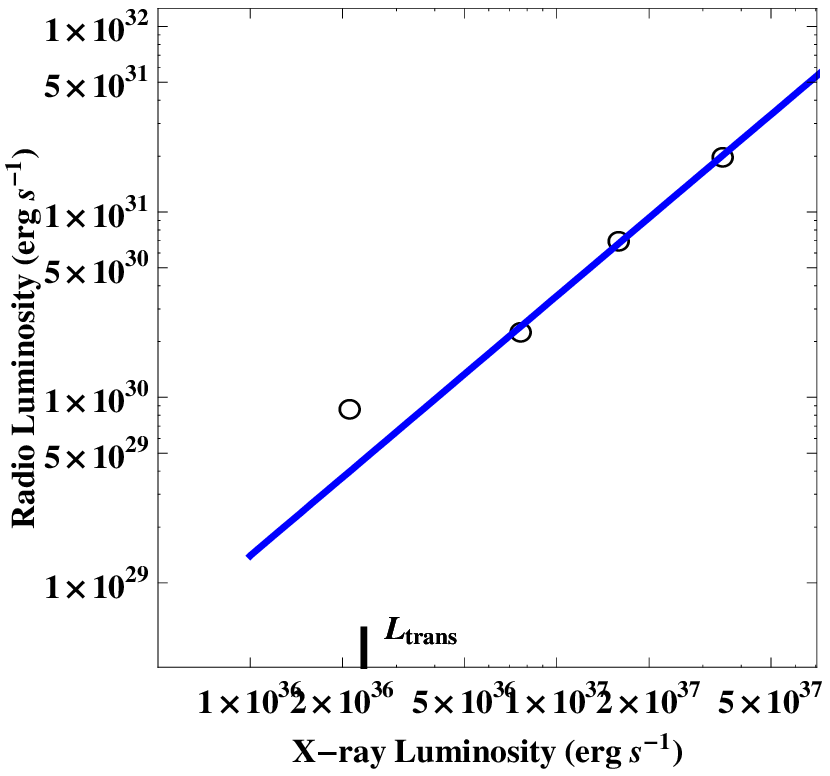} \hfill
\includegraphics[width=6.1cm]{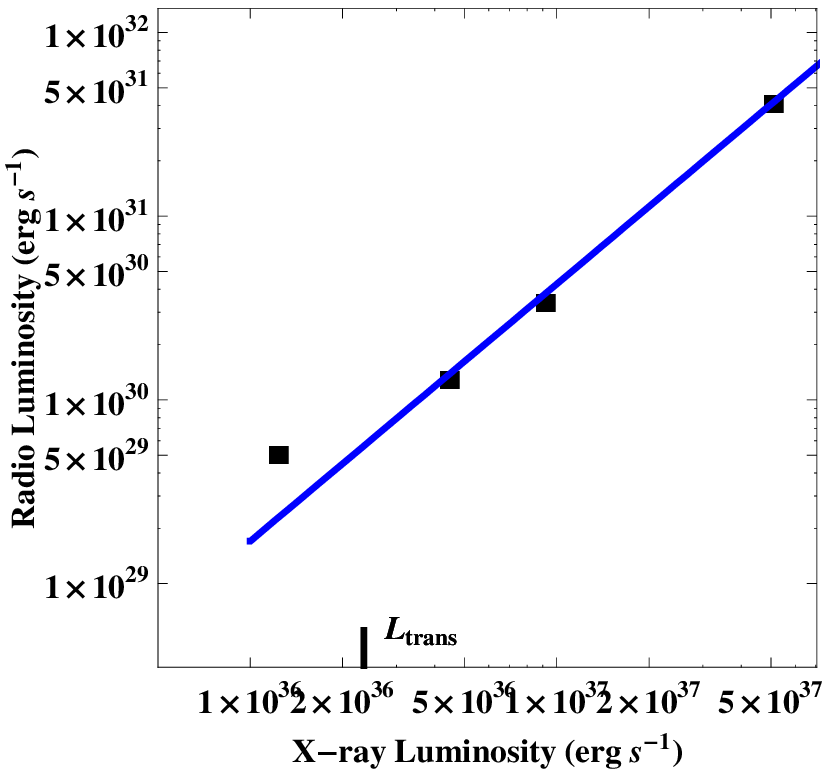}
\caption{The radio¨CX-ray correlations are plotted to fit the observation data based on our model for H1743-322 (left panel) and GX 339-4 (right panel). The four circles and squares are obtained based on the observations of the rising phase of LHS for H1743-322 and GX 339-4 given in Tables 1 and 2, respectively. The blue line corresponds to the correlation with an index of $\sim 1.4$} \label{fig4}
\end{center}
\end{figure}

\noindent Combining equations (12) and (16) we have the ratio of the BZ to BP powers as follows,


\begin{equation}
\label{eq18} P_{BP} / P_{BZ}^{max} = (2.56 \times 10^{-2})(f_0 / a_{*})^2 [(R_{tr} / R_H) - 1.95].
\end{equation}

Inspecting Fig. 1 and equations (10) and (17), we find that the ratio of the BZ to BP powers is sensitive to the truncated radius Rtr for the fixed BH spin. The contribution to the jet powerfrom the BZ power is roughly equal to that from the BP power, provided that the BH spin is no less that 0.1 with $R_{tr}$ greater than $100R_H$. However, the BZ power dominates over the BP power as $R_{tr}$ approaches to the BH for $R_{tr} /R_H \sim 3$. Since the BH spin remains almost unchanged in one outburst of BHXBs, these results imply that the evolution of the magnetic field configuration does play an important role in state transition from LHS to IMS of BHXBs. We shall discuss this issues in Section 5.

In order to fit the radio--X-ray correlation we should discuss the contribution of the jet power to radio luminosity and X-ray luminosity. Following HWW14 we have the radio luminosity related to the jet power by


\begin{equation}
\label{eq19} L_R = 6.1 \times 10^{-23} L_J^{17/12} erg \cdot s^{-1},
\end{equation}

\noindent where $L_J$ can be calculated by combining equations (11)--(17). Equation (19) is derived by Heinz {\&} Grimm (2005) based on the estimation of jet power for Cyg X-1 and GRS 1915 + 105.

Combining the spectra of LHS and equation (19) with the values of the input parameters listed in Table 3, we can fit the radio--X-ray correlation for H1743-322 and GX 339-4 of a steep power-law index $\sim 1.4$ as shown in the left and right panels of Fig. 4, respectively.

Considering that part of the jet power could be converted to X-ray radiation by some magnetic process, we assume that the radio radiation  $L_R$ is contributed by a fraction of the BZ and BP powers, and the X-ray radiation $L_X$ arises from the sum of the rest jet power and accretion flow, i.e.,


\begin{equation}
\label{eq20} L_X = L_X^{I} + L_X^{II},
\end{equation}

\noindent where  $L_X^{II}$ is contributed by accretion flow, and $L_X^{I}$ is related to the jet power as follows,


\begin{equation}
\label{eq21} L_X^{I} = \eta_i L_{J},
\end{equation}

The parameter  $\eta_i(i=1, 2, 3, 4)$ in equation (21) are the fraction of the jet power converted to X-ray luminosity corresponds to different observations. It is assumed that the jet contribution is distribute uniformly in the whole X-ray band, and $\eta_i$ can be determined in fitting the spectra of LHS with the associated $L_R-L_X$ correlations based on the four observation data of H1743-322 and GX 339-4 as shown in Table 4. It turns out that the steep radio-X-ray correlations can be fitted for H1743-322 and GX 339-4 as shown in Fig. 4.

As shown in Fig. 4 we have two kinds of radio--X-ray correlations corresponding to (i) from date 1 to date 2 with slopes $\sim 0.64$ for H1743-322 and $\sim 0.63$ for GX 339-4; (ii) from date 2 to date 4 with slope of $\sim 1.4$. The quantity $L_{trans}$ in Fig. 4 is the upper bound of the transition between the two correlations, being equal to $2.3 \times 10^{36} erg \cdot s^{-1}$. These results are consistent with those given by Coriat (2011).

\section{DISCUSSION}
\label{sect:discussion}

In this paper, we propose a model to interpret the rising phase of LHS of BHXBs based on the transportation of the magnetic field by accreting plasma from the companions. It turns out that the LHS spectra of H1743-322 and GX 339-4 can be fitted separately based on the four different observation data. In addition, the steep radio/X-ray correlations of these sources are fitted based on our model. Some issues related the role of the magnetic fields in state transitions of BHXBs are discussed as follow.

(i) Accretion disk is an ideal site for anchoring magnetic field. The transportation and amplification of magnetic field are related closely to the accreting plasma from companions of BHXBs, and it provides a reasonable picture for the origin of magnetic field in BHXBs, and is successful in explaining the main characteristics of LHS of H1743-322 and GX 339-4 in this model. It turns out that the main characteristics of LHS of H1743-322 and GX 339-4 could be interpreted based on this model. As shown in Tables 1 and 2, the truncated radius moves inwards with the transportation of the magnetic field in LHS. Thus we think that a stage for Magnetically Arrested Disk (MAD) might be a natural outcome of concentrating magnetic field in inner ADAF (Narayan, Igumenshchev {\&} Abramowicz 2003), giving rise to relativistic transient jets in IMS of BHXBs (Narayan {\&} McClintock 2012; McKinney, Tchekhovskoy {\&} Blandford 2012).

(ii) The state transition of BHXBs is governed by the evolution of the magnetic field configuration. The transition from LHS to HSS can be interpreted by a global magnetic field inversion in the MAD state as argued by Dexter et al. (2014), or by the conversion from poloidal dominated configuration to toroidal dominated one as pointed out by King et al. (2012).

(iii) Inspecting equation (20), we find that the transition X-ray luminosity consists of the contribution from accretion flow ($L_X^{II}$) and that from jet ($L_X^{I}$). Thus the quite different luminosity during different outbursts of the same source could be fitted based on equations (5) and (6), provided that $\dot{m}_0$ and $B_0$ are assigned to different values with the duration time $\tau$ for different outbursts.

(iv) The relation between the transition X-ray luminosity and rate-of-change of accretion rate could be derived based on our model, which has been found by Yu {\&} Yan (2009). Differentiating equation (5), we have

\begin{equation}
\label{eq22} (d \dot{m}/ dt)_{_{R_{tr}}} = \alpha_m \dot{m}_0 \tau^{-1} (t / \tau)^{\alpha_m - 1},
\end{equation}

\noindent where equation (22) is valid for the whole rising phase of LHS. Furthermore, we have accretion rate related to its rate-of-change by substituting $t = \tau$ into equation (22), i.e.,


\begin{equation}
\label{eq23} (d \dot{m}/ dt)_{_{R_{tr}}} = \alpha_m \dot{m}_0 \tau^{-1},
\end{equation}

It is noted that equation (23) is valid for the transition X-ray luminosity corresponding to IMS, and we can obtain a relation between this luminosity and rate-of-change of accretion rate by adjusting the concerning parameters $\alpha_m$, $\dot{m}_0$ and $\tau$. Incorporating equations (5) and (23), we find that the transition luminosity depends not only on the accretion rate, but also on its rate-of-change. This result is consistent with those given by Yu {\&} Yan (2009).

As to the exception occurred in Cyg X-1, we follow the explanation given in ZCH97: Cyg X-1 probably reflects a change in the relative importance of the energy release in the outer thin disk and the inner ADAF near the BH, and this may not require a substantial change in the total accretion rate, and our magnetic model is not applicable to this case.

Frankly speaking, there are a number of problems in this simplified model. First, since the origin and transportation of magnetic fields in black hole accretion remains unclear, we suggest that the magnetic flux is carried by the accreting plasma from companion, being governed by equations (5) and (6). Second, also due to lack of the detailed physical mechanisms of the origin of magnetic fields we have to adjust some parameters, such as $\alpha_m$ and $\alpha_{_B}$, to fit the main features in the rising phase of LHS of these two sources. Third, only the rising phase of LHS of the two BHXBs, H1743-322 and GX 339-4, is fitted in this paper, while the whole the time evolution from quiescent state to LHS, and to IMS is not discussed at all. We shall investigate these issues in our future work.

\begin{acknowledgements}
This work is supported by the National Basic Research Program of China (2009CB824800) and the National Natural Science Foundation of China (grants 11173011 and 11403003). We are very grateful to the anonymous referee for his (her) helpful suggestion for improving our work.

\end{acknowledgements}


\begin{thebibliography}{99}


\bibitem{1}{ Allured R., Tomsick J. A., Kaaret P., Yamaoka K., 2013, ApJ, 774, 135}


\bibitem{2}{Belloni T. M., Homan J., Casella
p., et al. 2005, A{\&}A, 440, 207}


\bibitem{2}{Blandford R. D., Znajek R. L., 1977, MNRAS, 179, 433}

\bibitem{3}{Blandford R. D., Payne D. G., 1982, MNRAS, 199, 883}

\bibitem{4}{Blandford R.D., Begelman M.C., 1999, MNRAS, 303, L1}

\bibitem{5}{Blum J. L., Miller J.M., Cackett E., et al. 2010, ApJ, 713, 1244}

\bibitem{6}{Cabanac C., Fender R. P., Dunn R. J. H., Kording E. G., 2009, MNRAS, 396, 1415}

\bibitem{7}{CadolleBel M., Ribo M., Rodriguez J., et al., 2007, ApJ, 659, 549)}

\bibitem{8}{Cao X. W., 2011, ApJ, 737, 94}

\bibitem{9}{Cao X. F., Wu Q. W., Dong A. J., 2014, ApJ, 788, 52}

\bibitem{10}{Corbel S., Fender R. P., 2002, ApJ, 573, L35}

\bibitem{11}{Corbel S., Nowak M. A., Fender R. P., et al., 2003, A{\&}A, 400, 1007}

\bibitem{12}{Corbel S., Fender R. P., Tomsick J. A., et al., 2004, ApJ, 617, 1272}

\bibitem{13}{Corbel S., Coriat M., Brocksopp C., et al., 2013, MNRAS, 428, 2500}

\bibitem{14}{Coriat M., Corbel S., Prat L., et al., 2011, MNRAS, 414, 677}

\bibitem{15}{Cowley A. P., Schmidtke P. C., Hutchings J. B., Crampton D., 2002, AJ, 123, 1741}

\bibitem{16}{Dexter J., McKinney, J. C., Markoff, S. {\&} Tchekhovskoy, A., 2014, MNRAS, 440, 2185}

\bibitem{17}{Doeleman S. S. et al. 2012, Science, 338, 355}

\bibitem{18}{Done C., Gierli¡änski M., Kubota A. 2007, A{\&}ARv, 15, 1}

\bibitem{19}{Done C., Diaz Trigo M., 2010, MNRAS, 407, 2287}

\bibitem{20}{Esin A. A., McClintock J. E., Narayan R. 1997, ApJ, 489, 865}

\bibitem{21}{Esin A. A., Narayan R., Cui W., Grove J. E., Zhang S. N., 1998, ApJ, 505, 854}

\bibitem{22}{Esin A. A., McClintock J. E., Drake J. J., et al. 2001, ApJ, 555, 483}

\bibitem{23}{Fender R. P., Gallo E., Jonker., 2003, MNRAS, 343, L99}

\bibitem{24}{Fender R. P., Belloni T. M., Gallo E., 2004, MNRAS, 355, 1105}

\bibitem{25}{Fender R. P., Russell D. M., Knigge C., et al. 2009, MNRAS, 393, 1608}

\bibitem{26}{Fender R. P., Belloni T. M., 2012, Science, 337, 540}

\bibitem{27}{Gallo E., Fender R. P., Pooley G. G., 2003, MNRAS, 344, 60}


\bibitem{28}{Gallo E., Corbel S., Fender R. P., Maccarone T. J., Tzioumis A. K., 2004, MNRAS, 347, L52}
\bibitem{29}{Guilet, J., Ogilvie, G. I., 2013, MNRAS, 430, 822}

\bibitem{30}{Ghosh P., Abramowicz M. A., 1997, MNRAS 292, 887}
\bibitem{31}{Hannikainen D. C., Hunstead R.-W., Campbell-Wilson D., Sood R. K., 1998, A{\&}A, 337, 460}
\bibitem{32}{Hawley J. F., Balbus S. A. 2002, ApJ, 573, 738}
\bibitem{33}{Heinz S., Grimm H. J., 2005, ApJ, 633, 384}
\bibitem{34}{Homan J., Wijnands R., van der Klis M., et al. 2001, ApJS, 132, 377}
\bibitem{35}{Homan J., Belloni T. M., 2005, ApSS, 300, 107}
\bibitem{36}{Huang C. Y., Wu Q. W., Wang D. X., 2014, MNRAS, 440, 965(HWW14)}
\bibitem{37}{Hynes R. I., Steeghs D., Casares J., Charles P. A., O¡¯Brien K., 2003, ApJ, 583, L95}
\bibitem{38}{Hynes R. I., Steeghs D., Casares J., Charles P. A., O¡¯Brien K., 2004, ApJ, 609, 317}
\bibitem{39}{Igumenshchev, I. V. 2009, ApJ, 702, L72}
\bibitem{40}{Igumenshchev, G. V., Narayan R., Abramowicz M. A., 2003, ApJ, 592, 1042}
\bibitem{41}{Joinet A., Jourdain E., Malzac J., et al. 2005, ApJ, 629, 1008}
\bibitem{42}{Jonker P. G., Miller-Jones J., Homan J., et al. 2010, MNRAS, 401, 1255}
\bibitem{43}{King A. L., Miller J. M., Raymond J., et al. 2012, ApJ, 746, L20}

\bibitem{44}{Kylafis , Belloni T. M., 2015, A{\&}A, 574, 133}

\bibitem{45}{Lei W. H., Wang D. X., Ma R. Y., 2005, ApJ, 619, 420}
\bibitem{46}{Lei W. H., Wang D. X., Zou Y. C., Zhang L., 2008, ChJAA, 8, 404}
\bibitem{47}{Livio M., Ogilvie G. I., Pringle J. E., ApJ, 1999, 512, 100}
\bibitem{48}{Livio M., 2002, Nature, 417, 125}
\bibitem{49}{Li Y., Gan Z. M., Wang D. X., 2010, New Astronomy, 15, 102}
\bibitem{50}{Lubow, S. H., Papaloizou, J. C. B., Pringle, J. E. 1994, MNRAS, 267, 235}
\bibitem{51}{Macdonald D., Thorne K. S., 1982, MNRAS, 198, 345}
\bibitem{52}{McClintock J. E., Remillard R. A. 2006, In Compact Stellar X-ray Sources, ed. WHG Lewin, M, van der Klis, pp. 157¨C214. Cambridge: Cambridge University Press}
\bibitem{53}{McClintock J. E., Remillard R. A., Rupen M. P., et al. 2009, ApJ, 698, 1398}
\bibitem{54}{McKinney J. C., Tchekhovskoy A., Blandford R. D., 2012, MNRAS, 423, 3083}
\bibitem{55}{Miller J. M., Fabian A. C., Reynolds C. S., et al. 2004, ApJ, 606, L131}
\bibitem{56}{Miller J. M., Homan J., Steeghs D., et al. 2006, ApJ, 653, 525}
\bibitem{57}{Miller J. M., Reynolds C. S., Fabian A. C., Miniutti G., Gallo L. C., 2009, ApJ, 697, 900}
\bibitem{58}{Miller J. M., Reynolds C. S., Fabian A. C., et al. 2012, ApJ, 759, L6}
\bibitem{59}{Miyamoto S., Kitamoto S., Hayashida K., Egoshi W., 1995, ApJL, 442, 13}


\bibitem{60}{Motta S., Munoz-Darias T., Belloni T., 2011, MNRAS, 408, 1796}
\bibitem{61}{Munoz-Darias T., Casares J., Mart¡ä?nez-PaisI. G., 2008, MNRAS, 385, 2205}
\bibitem{62}{Nandi A., Debnath D., Mandal S., Chakrabarti S. K., 2012, A{\&}A, 542, A56}
\bibitem{63}{Narayan R., YiI., 1994, ApJ, 428, L13}
\bibitem{64}{Narayan R., YiI., 1995, ApJ, 444, 231}
\bibitem{65}{Narayan R., Igumenshchev I.V., Abramowicz M. 2000, ApJ, 539, 798}
\bibitem{66}{Narayan R., Igumenshchev I. V., Abramowicz M. A., 2003, PASJ, 55, L69}
\bibitem{67}{Narayan R., McClintock J. E., 2012, MNRAS, 419, 69}
\bibitem{68}{Paczynski B., 1971, ARA{\&}A, 9, 183}
\bibitem{69}{Plant D. S., Fender R. P., Ponti G., Munoz-Darias T., Coriat M., 2014, MNRAS, 442, 1767}
\bibitem{70}{Quataert E., Gruzinov A., 2000, ApJ, 539, 809}
\bibitem{71}{Ratti E. M., Jonker P. G., Miller-Jones J. A. C., et al. 2012, MNRAS, 423, 2656}
\bibitem{72}{Reis R. C., Fabian A. C., Ross R. R., Miniutti G., Miller J. M., Reynolds C., 2008, MNRAS, 387, 1489}
\bibitem{73}{Remillard R. A., McClintock J. E., Orosz J. A., Levine A. M., 2006, ApJ, 637, 1002}
\bibitem{74}{Sikora, M., Begelman M. C., 2013, ApJ, 764, L24}
\bibitem{75}{Soleri P., Fender R., Tudose V., et al. 2010, MNRAS, 406, 1471}
\bibitem{76}{Spruit H. C., Uzdensky D. A., 2005, ApJ, 629, 960}
\bibitem{77}{Spruit H. C. 2010, The Jet Paradigm, Lecture Notes in Physics, Vol 794, Springer-Verlag Berlin Heidelberg, p. 233, arXiv:0804.3096}
\bibitem{78}{Sriram K., Agrawal V. K., Rao A. R., 2009, RAA, 9, 901}
\bibitem{79}{Stone J., Pringle J., Begelman M., 1999, MNRAS, 310, 1002}
\bibitem{80}{Tomsick J. A., Yamaoka K., Corbel S., et al. 2009, ApJ, 707, L87}

\bibitem{82}{Wu Q. W., Cao X. W., 2006, PASP, 118, 1098}
\bibitem{83}{Wu Q. W., Gu M. F., 2008, ApJ, 683, 212}
\bibitem{84}{Wu Q. W., Cao X. W., Wang D. X., 2011, ApJ, 735, 50}
\bibitem{85}{Wu Q. W., Yan H., Yi Z.,2013, MNRAS, 436, 1278}
\bibitem{86}{Yuan F., 1999, ApJ, 521, L55}
\bibitem{87}{Yuan F., Ma R.Y., Narayan R., 2008. ApJ, 679, 984 (YMN08)}
\bibitem{88}{Yuan F., 2001, MNRAS, 324, 119}
\bibitem{89}{Yuan F., Cui W., Narayan R., 2005, ApJ, 620, 905}
\bibitem{90}{Yuan F., Narayan R., 2014, ARAA, 52, 529}
\bibitem{91}{Yu W. F., van der Klis M., Fender R., 2004, ApJL, 611L, 121}
\bibitem{92}{Yu W. F., Yan Z., 2009, ApJ, 701, 1940}
\bibitem{93}{Zdziarski A. A., Gierlinski M., Mikolajewsk\emph{\emph{}}a J., et al. 2004, MNRAS, 351, 791}
\bibitem{94}{Zhang S. N., Cui W., Harmon B. A., et al. 1997, ApJ, 477, L95 (ZCH97)}
\end{thebibliography}
\end{document}